\documentclass[10pt]{elsart}

\usepackage{natbib}
\usepackage{graphicx}

\usepackage{amssymb}
\usepackage{amsxtra}

\begin{document}

\begin{frontmatter}


\title{On the existence and structure of a mush\\ at the inner core boundary of the Earth}


\author{R. Deguen\corauthref{cor1}},
\corauth[cor1]{Corresponding author.}
\ead{renaud.deguen@obs.ujf-grenoble.fr}
\author{T. Alboussi\`ere},
\author{D. Brito}

\address{Laboratoire de G\'eophysique Interne et Tectonophysique, Universit\'e Joseph Fourier, Grenoble, France }

\begin{abstract}

It has been suggested about 20 years ago that the liquid close to the inner core boundary (ICB) is supercooled and that
a sizable mushy layer has developed during the growth of the inner core.
The morphological instability of the liquid-solid interface which usually results in the formation of a mushy zone has been intensively studied in metallurgy, but the freezing of the inner core occurs in very unusual conditions: the growth rate is very small, and the pressure gradient has a key role, the newly formed solid being hotter than the adjacent liquid. 
 
We investigate the linear stability of a solidification front under such conditions, pointing out the destabilizing role of the thermal and solutal fields, and the stabilizing role of the pressure gradient. 
The main consequence of the very small solidification rate is the importance of advective transport of solute in liquid, which tends to remove light solute from the vicinity of the ICB and to suppress supercooling, thus acting against the destabilization of the solidification front. For plausible phase diagrams of the core mixture, we nevertheless found that the ICB is likely to be morphologically unstable, and that a mushy zone might have developed at the ICB. 
The thermodynamic thickness of the resulting mushy zone can be significant, from $\sim100$ km to the entire inner core radius, depending on the phase diagram of the core mixture.
However, such a thick mushy zone is predicted to collapse under its own weight, on a much smaller length scale ($\lesssim 1$ km).
We estimate that the interdendritic spacing is probably smaller than a few tens of meter, and possibly only a few meters.

\end{abstract}


\end{frontmatter}

\section{Introduction}

As the Earth's core is gradually cooling down, the inner core is freezing from the liquid core mixture \citep{Jacobs1953} thought to be mostly iron-nickel alloyed with a small quantity of lighter elements. 
Although the nature and relative abundances of these light elements are still controversial \citep{Poirier1994}, mineralogical models consistent with the seismological constraints indicate that their concentration is greater, perhaps by a factor 4, in the liquid core than in the inner core \citep{Anderson1994}: like most alloys, the core mixture is fractionating during the process of crystallization, the liquid core being slowly enriched in light elements.

The release of light elements which results from the gradual solidification of the core is thought to be a major source of buoyancy for driving the core convection and the geodynamo \citep{Braginsky1963,Gubbins04},
but segregation of light solute may also have dramatic consequences on the structure of the solid inner core itself \citep{Fearn1981}. 
It is well known from metallurgical experiments that solute segregation during solidification can result in the supercooling of the liquid close to the solidification front. 
This supercooling is usually suppressed by either solidification of isolated crystals in a slurry layer, or by the development of a mushy zone, where solid cells or dendrites coexist with a solute rich liquid \citep[e.g.][]{Kurz89}.

\citet{Loper1981} and \citet{Fearn1981} show that the conditions at the inner core boundary (ICB) are almost certainly favorable to the formation of either a slurry layer or a mushy layer. \citet{Fearn1981} gave preference to the latter and concluded that a mushy zone of considerable depth, possibly extending to the inner core center, must have developed at the ICB. More recently, \citet{Shimizu2005} quantitatively studied the possible regimes of solidification (slurry layer or mushy zone) and found that a dendritic regime is more probable than a slurry layer regime, because of the difficulty to supply enough nuclei to feed a slurry layer. Some seismological study \citep[e.g.][]{Cao2004} also argue in favor of the existence of a mushy zone at the top of the inner core.

\citet{Morse1986,Morse2002} challenged this view arguing that convective motion should quantitatively remove light solute from the inner core boundary, thus suppressing supercooling. 
From an analogy between the growth of the inner core and cumulate solidification in magma chambers, \cite{Morse2002} concluded that the ICB must be microscopically flat, and rejected the idea of a mushy zone at the top of the inner core.
Cumulates and dendritic layers have dynamics similar in many respects, both being reactive porous media, but their development occurs through two physically different processes. 
Cumulates form by sedimentation of early formed crystals from the melt, the possible porosity resulting from a competition between sedimentation of crystals and solidification in the layer.
In contrast, a dendritic layer develops as a consequence of an instability of an initially plane solid-liquid interface.
Whether Morse's analysis is appropriate or not to the formation of a dendritic layer may therefore be questionable, but it is probably correct that advective transport of solute may affect the solidification regime of the ICB.

The morphological stability of a solidification front has been intensively studied both theoretically and experimentally \citep[e.g.][]{MS1963,Kurz89,Davis01} but, as noted by \citet{Shimizu2005}, the freezing of the inner core occurs in very unusual conditions: the growth rate is very small, about 6 orders of magnitude smaller than in typical laboratory experiments, and the pressure gradient has a key role, the newly formed solid being hotter than the adjacent liquid.

In this work, we will take into account first order effects of advective transport and investigate possible effects of the very slow solidification rate and of the pressure gradient on the stability of a solidification front (sections 2, 3 and 4).
In section 5, the thermophysical parameters and conditions relative to the crystallization of the inner core will be evaluated from a survey of the literature and from the use of thermodynamic constraints.
The stability of the ICB will then be discussed and we shall argue that the existence of a mushy zone is very likely.
Finally, in section 6, we will estimate various length scales relevant to the structure of a mushy inner core, such as its depth and interdendritic spacing.

\section{Supercooling in the vicinity of an interface}

When a dilute alloy is frozen from the melt, the newly formed solid usually differs in composition from the liquid, in a way which depends on the solubility of the solute in the solid and liquid phases. This solute redistribution  is conveniently described by the distribution coefficient $k=c^{\,\mathrm{s}}/c$, where $c^{\,\mathrm{s}}$ and $c$ are respectively the mass fractions of solute in the solid and in the liquid.
In the common case where $k<1$, solidification results in the rejection of solute from the solid phase, the liquid phase being enriched in solute in the vicinity of the solidification front.

As the melting temperature is lowered by the presence of solute, the liquidus temperature at the interface is lower than that of the liquid ahead of the interface, which is richer in solute. If the actual temperature gradient at the solidification front is smaller than the gradient of the melting temperature, the temperature of the liquid close to the interface is smaller than the liquidus temperature, i.e. the liquid is supercooled. 
This non-equilibrium state usually does not persist and results in the formation of either a mushy zone, through a destabilization of the solidification front, or a slurry layer.

Denoting $T^{\ell}$ and  $T_{\mathrm{m}}$ the temperature in the liquid and the melting temperature respectively, the criterion for supercooling is:
\begin{equation}
\left.\frac{d T^{\ell}}{d z}\right|_0 <  \left.\frac{d T_{\mathrm{m}}}{d z}\right|_0
\end{equation}
where the subscript $0$ denotes the value at the solid-liquid interface, and where the $z$ axis points toward the liquid.
In the core, the melting temperature depends on both the solute concentration $c$ and the pressure $P$. Defining the liquidus and Clapeyron slopes as:
\begin{equation}
m_{\,\mathrm{c}} = \left.\frac{\partial T_{\,\mathrm{m}}}{\partial c}\right|_{\,\mathrm{P}}<0 \quad \hbox{and} \quad
m_{\,\mathrm{P}} = \left.\frac{\partial T_{\,\mathrm{m}}}{\partial P}\right|_{\,\mathrm{c}}>0,
\end{equation}
the supercooling criterion can be written:
\begin{equation}
G_{\mathrm T}^{\ell} - m_{\,\mathrm{P}}\, G_{\mathrm P} <  m_{\,\mathrm{c}}\, G_{\mathrm c}
\label{eq:crit}
\end{equation}
where $\displaystyle{G_{\mathrm T}^{\ell} = \left.\frac{d T^{\ell}}{d z} \right|_0}$, $\displaystyle{G_{\mathrm c} = \left.\frac{d c}{d z}\right|_0}$ and $\displaystyle{G_{\mathrm P} = \left.\frac{d P}{d z}\right|_0}$ are respectively the thermal, chemical and pressure gradients at the solid-liquid interface.
In the solidification conditions of the core, the pressure gradient is negative and $m_{\,\mathrm{P}}$ positive \citep[e.g.][]{Boehler1993}, thus $m_{\,\mathrm{P}} G_{\mathrm{P}}$ is negative and the pressure field acts against supercooling. 
In contrast, the temperature and concentration fields both promote supercooling, as $G_{\mathrm T}^{\ell}<0$ and $m_{\,\mathrm{c}}\, G_{\mathrm c}>0$.
We can therefore predict that the pressure field should have a stabilizing effect on the solidification front whereas temperature and solutal fields should be destabilizing.

\section{The effect of advective transport on the supercooling criterion \label{sec:adv}}

If motion is present in the liquid, advective transport of solute, and, to a smaller extent, of heat, may affect the solutal and thermal profiles near the solid-liquid boundary, and thus the degree of supercooling.
In this section, we consider the effects of buoyancy-driven convection on the mean solute and thermal fields. 
Let us consider the directional solidification of a dilute binary alloy at constant velocity $V$. 
The solute has a concentration $c_{\infty}$ in the bulk of the liquid and a chemical diffusivity $D_c$ in the liquid. The coordinate system is fixed on the moving front, the $z$ axis pointing toward the liquid perpendicularly to the solidification front. The liquid has a velocity $\mathbf u$.

In the melt, the solute concentration $c$ is given by the equation of conservation in the bulk:
\begin{subequations}
\begin{equation}
-V\frac{\partial c }{\partial z}+(\mathbf u \cdot \boldsymbol{\nabla} ) c = D_c\ \nabla^2 c  , 
\label{eq:c}
\end{equation}
with boundary conditions at infinity:
\begin{equation}
c \rightarrow c_{\infty} \quad \hbox{at} \quad z \rightarrow \infty ,
\end{equation}
and at the interface $z=0$, given by the equation of conservation of solute at the solidification front:
\begin{equation}
G_c= (c^s_0-c_0 ) \frac{V}{D_c} = c_0 (k-1) \frac{V}{D_c},
\label{eq:c_I}
\end{equation}
\end{subequations}
where $c^s_0$ and $c_0$ are the concentration in the solid and liquid phases at the interface.
If advective transport is negligible, a solute boundary layer of thickness $\delta_c=D_c/V$ will build up, and after a transient stage during which the solute concentration in the solid at the solidification front gradually increases from $k c_{\infty}$ to $c_{\infty}$, the system reaches a steady state where the concentration of solute $c(z)$ in the liquid is
\begin{equation}
c(z) = c_{\infty}  + c_{\infty}\frac{1-k}{k}\exp \left(-z/\delta_c\right)
\label{eq:c_diff}
\end{equation}
\citep[e.g.][]{Kurz89,Davis01}.
The concentration in the liquid at the interface is $c_0 = c_{\infty}/k$ and the chemical gradient at the interface is
\begin{equation}
G_c= c_\infty \frac{k-1}{k} \frac{V}{D_c}.
\label{Gc_diffusif}
\end{equation}
However, if the solidification is very slow, as it is at the ICB, advection may affect strongly the concentration gradient at the interface. With a solidification velocity of $10^{-11}\ \mathrm{m.s^{-1}}$ and a chemical diffusivity of order $10^{-9} \ \mathrm{m^2.s^{-1}}$, the chemical diffusive boundary layer thickness is $\delta_c = D_c/V \sim 100\ \mathrm{m}$.
Defining the solutal Rayleigh number $\mathrm{Ra^s}$ as:
\begin{equation}
\mathrm{Ra^s}=\frac{\beta g G_c \delta_c^4}{\nu D_c}, 
\end{equation}
where $\beta\sim 1$ \citep{Gubbins04} is the compositional expansion coefficient, $g=4.4\ \mathrm{m.s}^{-2}$ is the magnitude of gravity at the ICB \citep{PREM} and $\nu \sim 10^{-6}\ {\mathrm m^2 s^{-1}}$ is the kinematic viscosity of the liquid, we found $\mathrm{Ra^s} \sim 10^{21} $. This is much larger than the critical Rayleigh number for the Rayleigh-Taylor instability.
The purely diffusive boundary layer is therefore subject to convective instabilities, which remove light solute from the vicinity of the interface much more efficiently than diffusion alone.

The purely diffusive concentration profile described by equation \eqref{eq:c_diff} is therefore not relevant to the case of the Earth's inner core solidification, and effects of advective solute transport on the mean concentration profile must be investigated.
In the case of purely diffusive solute transport, $c_0$ is equal to $c_{\infty}/{k}$. This is the maximum allowed value as a greater value will imply a solute concentration in the solid greater than $c_{\infty}$. 
If convective motions are present, advective transport results in the depletion of solute close to the solidification front, and the concentration in the liquid at the boundary will be closer to $c_{\infty}$ as convection is more vigorous, whereas concentration in the solid will tend to $k c_{\infty}$.
Because solute conservation implies that the concentration in the solid cannot exceed $c_{\infty}$ and must be greater than $k c_{\infty}$, the concentration gradient $G_{\mathrm c}$ at the interface is bounded by:
\begin{equation}
c_{\infty}(k-1)\frac{V}{D_c} \leq G_{\mathrm{c}} \leq c_{\infty} \frac{k-1}{k}\frac{V}{D_c} ,
\label{eq:bounds}
\end{equation}
the upper bound being the value of the chemical gradient in the purely diffusive case. This gradient tends toward its lower bound as the efficiency of advective transport is enhanced; in what follows, we will consider that the actual chemical gradient can be approximated by its lower bound: 
\begin{equation}
G_{\mathrm{c}} \simeq c_{\infty}(k-1)\frac{V}{D_c} ,
\label{eq:lowerbound}
\end{equation}
which should be compared with equation \eqref{Gc_diffusif}.
If $k$ is small, the convective chemical gradient can be considerably smaller than the diffusive gradient.

With a no-slip boundary condition, $\mathbf u(x,y,0) = \mathbf 0$ for all $(x,y)$, equation \eqref{eq:c} taken at $z=0$ gives the second derivative of the mean concentration profile at the interface:
\begin{equation}
\left. \frac{d^2  c}{d z^2} \right|_0 = -\left.\frac{V}{D_c} \frac{d c}{d z} \right|_0 = - \frac{1}{\delta_c} G_{\mathrm{c}} .
\end{equation}
In what follows, the vertical variations of the mean concentration field in the vicinity of the interface will be described by a second order Taylor expansion:
\begin{equation}
c(z) = c_0 + G_{\mathrm{c}} z - G_{\mathrm{c}} \frac{1}{\delta_c} \frac{z^2}{2}.
\label{eq:c_m}
\end{equation}

Because the thermal diffusivity is much larger than the solute diffusivity $D_c$, advection affects the solute flow much more strongly than the thermal flow, and we will ignore the effects of advection on the thermal field, i.e. the mean temperature is solution of the equation of diffusion in the liquid and in the solid:
\begin{equation}
-V\frac{\partial T^{\ell}}{\partial z}=
D_{\mathrm T}^{\ell} \nabla^2 T^{\ell} ,
\end{equation}
\begin{equation}
-V\frac{\partial T^{\mathrm s}}{\partial z} 
=D_{\mathrm T}^{\mathrm s} \nabla^2 T^{\mathrm s }, 
\end{equation}
where $D_{\mathrm T}^{\ell}$ and $D_{\mathrm T}^s$ are the thermal diffusivities in the liquid and solid respectively.
The thermal gradient at the interface is given by the heat balance at the solid-liquid interface: 
\begin{equation}
L_v  V =  \left( \kappa^s G_{\mathrm{T}}^s - \kappa^{\ell} G_{\mathrm{T}}^{\ell}  \right) ,
\label{eq:heat_balance}
\end{equation}
where $L_v$ is latent heat per unit volume, $\kappa^s$ and $\kappa^{\ell}$ are the thermal conductivities in the solid and liquid respectively and $G_{\mathrm{T}}^s$ and $G_{\mathrm{T}}^{\ell}$ are the gradient of temperature at the interface in the solid and liquid respectively.
The temperature field can be expanded as
\begin{equation}
{T}(z) = T_0 + G_{\mathrm{T}}^{\ell} z - G_{\mathrm{T}}^{\ell} \frac{1}{\delta_T} \frac{z^2}{2},
\label{eq:T_m}
\end{equation}
where $\delta_T = D_{\mathrm T}/V$.

\citet{Shimizu2005} did not take into account effects of convection on the thickness of the boundary layer, and assumed it to be given by the diffusive boundary layer thickness $\delta_c = D_c/V$.
They study the possibility that a fraction of the solidification occurs in the supercooled zone, whose thickness has been taken to be the diffusive boundary layer thickness, that is $\sim 100$ m.
With this value, they found that it is unlikely that a significant fraction of the solidification occurs in a slurry layer, because of the difficulty of supplying continuously enough nuclei.
Taking into account advective transport leads to a much thinner supercooled zone, with smaller supercooling, where the amount of crystals solidified is probably much smaller than expected from the model of \citet{Shimizu2005}.
We will therefore consider that all the solidification occurs at the solid-liquid interface, and that growth of crystals in the supercooled zone is negligible.

\section{Linear stability analysis of the growth interface of a binary alloy in a pressure gradient \label{section4}}

\subsection{Solutal and thermal instabilities}

Considerations of section \ref{sec:adv} allow us to determine a mean state for the solute and thermal fields, hereafter noted $\bar c$ and $\bar T$, respectively described by equations \eqref{eq:c_m} and \eqref{eq:T_m}; the pressure field is taken to be hydrostatic, the pressure gradient being $G_{\mathrm P}^{\ell} = - \rho^{\ell} g$.

In this section we consider the stability of the planar interface against infinitesimal perturbations of the mean fields. The interface is not planar anymore, but has an infinitesimal topography $h(x,y,t)$ which temporal evolution we study. 
The solute and thermal fields can be expressed as the sum of the mean field and infinitesimal disturbances : $c(x,y,z,t) = \bar c (z) + \tilde c (x,y,z,t)$ and $T(x,y,z,t) = \bar T (z) + \tilde T (x,y,z,t)$.
The linear stability analysis we propose here is similar in principle to Mullins and Sekerka's analysis \citep{MS1963}. 
The two main differences concern the formulation of the basic state, which is considered here to be altered by convective motions, and the dependence of the melting temperature on pressure.

The thermal and solutal fields must satisfy:

(i) the equations of conservation of solute and heat in the liquid and solid phases:
\begin{subequations}
\begin{equation}
\frac{\partial c }{\partial t}-V\frac{\partial c}{\partial z}
=D_c\ \nabla^2 c  ,
\end{equation}
\begin{equation}
\frac{\partial T^{\ell}}{\partial t}-V\frac{\partial T^{\ell}}{\partial z}=
D_{\mathrm T}^{\ell} \nabla^2 T^{\ell} ,
\end{equation}
\begin{equation}
\frac{\partial T^{\mathrm s}}{\partial t}-V\frac{\partial T^{\mathrm s}}{\partial z} 
=D_{\mathrm T}^{\mathrm s} \nabla^2 T^{\mathrm s}  .
\end{equation}
Diffusion of solute in the solid is neglected.

(ii) the boundary conditions at the interface.
At thermodynamic equilibrium, the temperature at the interface $T^I$ must obey the Gibbs-Thomson relation:
\begin{equation}
T^I = T_{\mathrm m} + \Gamma H + m_{\mathrm P} G_{\mathrm P}^{\ell} h +m_{\mathrm c} c ,
\end{equation}
which states that the temperature at the interface is equal to the melting temperature of the curved interface, when variations with pressure, concentration and interface curvature are taken into account. $T_{\mathrm m}$ is the solvent melting temperature of a flat interface at $z=0$,  $H\simeq \nabla ^2 h $ is the interface curvature and $\Gamma = T_{\mathrm m} \gamma / L_v$ is the Gibbs coefficient, where $\gamma$ is the liquid-solid interfacial energy.

In addition, the thermal and solutal fields must respectively satisfy the heat balance and the solute conservation at the interface:
\begin{equation}
L_v  v_{n}(x,y) =  \left( \kappa^{\mathrm s}\boldsymbol{\nabla} T^{\mathrm s} -  \kappa^{\ell}\boldsymbol{\nabla} 
T^{\ell} \right)\cdot\boldsymbol{n}(x,y),
\end{equation}
\begin{equation}
\left(c^{\mathrm s}-c\right) v_{n}(x,y)=D_c\ \boldsymbol{\nabla} c \cdot\boldsymbol{n}(x,y) ,
\end{equation}
\label{eq:syst}
\end{subequations}
where $v_{n}(x,y) \simeq V+\partial h / \partial t$ is the speed of the front and $\boldsymbol{n}$ is the unit normal vector pointing toward the liquid.

(iii) boundary conditions at infinity: the perturbations must decay to zero at infinity (but see discussion below).

Equations \eqref{eq:syst} lead to a system in perturbation quantities $h$, $\tilde c$, $\tilde T^{\ell}$ and $\tilde T^s$ which is linearized, allowing to seek solutions for the perturbed fields under the following normal mode form:
\begin{equation}
\begin{pmatrix} h \\ \tilde{c} \\ \tilde{T}^{\ell} \\ \tilde{T}^{s} \end{pmatrix}
= 
\begin{pmatrix} h_1 \\ c_1(z) \\ T^{\ell}_1(z) \\ T^s_1(z) \end{pmatrix}
\exp(\omega t + \mathrm{i} k_x x + \mathrm{i} k_y y)
\label{perturbations}
\end{equation}
where $\omega$ is the growth rate of the disturbance and $k_x$ and $k_y$ are the wave numbers along the interface in the direction $x$ and $y$. 
From equations (\ref{eq:syst}a,b,c) and the boundary condition (iii), $c_1(z)$, $T^{\ell}_1(z)$ and  $T^s_1(z)$ are found to be proportional to $\exp(\beta z)$, $\exp(\beta^\ell z)$ and $\exp(\beta^s z)$ respectively, where $\beta$, $\beta^\ell$ and $\beta^s$ are
\begin{gather*}
\beta^{\mathrm s} = -\frac{1}{2} \frac{V}{D_\mathrm{T}^s} \left[ 1 - \sqrt{1+4\left(\frac{D_{\mathrm T}^{\mathrm s} k_h}{V} \right)^2} \right] \geq 0, \\
\beta^{\ell} = -\frac{1}{2} \frac{V}{D_\mathrm{T}^\ell} \left[ 1 + \sqrt{1+4\left(\frac{D_{\mathrm T}^{\ell} k_h}{V}\right)^2} \right] \leq -\frac{V}{D_\mathrm{T}^\ell},\\
\beta = -\frac{1}{2} \frac{V}{D_c}\left[ 1 + \sqrt{1+4\left( \frac{D_c k_h}{V} \right)^2} \right] \leq -\frac{V}{D_c}.\\
\end{gather*}
Using expression \eqref{perturbations} in the perturbation equations leads to a set of homogeneous, linear equations in $h_1$, $c_1$, $T_1^{\ell}$ and $T_1^s$ which have nontrivial solutions if the following dispersion equation is satisfied:
\begin{equation}
\begin{split}
\omega =&
\Biggl\{
\alpha m_{\mathrm{c}}G_{\mathrm{c}}
-\alpha^{\ell} G_{\mathrm T}^{\ell}
-\alpha^s G_{\mathrm T}^{\mathrm s} 
+m_{\mathrm{P}}G_{\mathrm{P}}
-\Gamma k_{h}^2
\Biggr\}  \\
&\Bigg/ \Biggl\{
\dfrac{L_v }{\kappa^s \beta^{\mathrm s} - \kappa^{\ell} \beta^{\ell}}-\frac{ D_c/V^2 }{\beta D_c/V+1-k}m_{\mathrm{c}}G_{\mathrm{c}}
\Biggr\} ,
\label{eq:dispersion}
\end{split} 
\end{equation}
where:
\begin{align*}
 k_h &= \sqrt{k_x^2 + k_y^2}, \\
\alpha^{\ell}&=\frac{\kappa^\ell \beta^{\ell}+\kappa^\ell V/D_\mathrm{T}^\ell}{\kappa^\ell \beta^{\ell}-\kappa^\mathrm{s}\beta^s} \in \left[0\ \frac{\kappa^\ell}{\kappa^\ell+\kappa^s}\right]  ,\\
\alpha^{\mathrm s} &= -\frac{\kappa^s \beta^s+\kappa^s V/D_\mathrm{T}^s}{\kappa^\ell \beta^{\ell}-\kappa^\mathrm{s}\beta^s} \in \left[\frac{\kappa^s}{\kappa^\ell+\kappa^s}\ \frac{\kappa^s D_\mathrm{T}^\ell}{\kappa^\ell D_\mathrm{T}^s}\right], \\
\alpha &=  \frac{\beta+V/D_c}{\beta+(1-k)V/D_c} \in [0\ 1] .\\
\end{align*}

Stability of the interface against infinitesimal perturbations depends on the sign of $\omega$: the solidification  front is stable if $\omega$ is negative for all wave numbers whereas it is unstable if $\omega$ is positive for any wave number.
As $\kappa^s\beta^{\mathrm s}-\kappa^\ell\beta^{\ell} \geq 0$, $\beta D_c/V+1-k \leq 0$ and $m_{\mathrm{c}}G_{\mathrm{c}} > 0$, the denominator is always positive and so the sign of $\omega$ depends only on the sign of the numerator, which is a combination of temperature, concentration and pressure gradient weighted by wave number dependent functions.

Convection has been taken into account when estimating the mean chemical gradient, but to be rigorous, we should have considered the effects of convection on the perturbed fields as well.
We might have imposed, as \citet{Coriell76} or \citet{Favier83} did, that perturbations of the solute field must decay to zero on a finite length, imposed by convection, but numerical calculations not presented here show that this does not significantly affect the prediction of the stability of the interface. 
This can be understood as follows. With the additional hypothesis of a small solutal Peclet number $V/k_h D_c$ (that is, for instability wavelengths small compared to $2 \pi D_c/V \simeq 300$ m, an hypothesis which will be shown  in section 5 to be well verified), $\beta$ is very close to $-k_h$, and $\alpha$ is very close to one because $k_h \gg V/D_c$.
If taken into account, advection of the perturbed solutal field would decrease the decay length of the perturbations, and so increase $\beta$. This would make $\alpha$ to be even closer to one, but will hardly affect the value of the chemical term in the numerator. Effects on the stability limit and on the critical wavelength would therefore be insignificant. 
However, an increase of $\beta$ would result in a decrease of the denominator, and hence in an increase of $\omega$: the instability growth rate may be underestimated by equation \eqref{eq:dispersion}, but this point will not weaken our analysis of the ICB solidification regime.

The dispersion relation confirms the qualitative differences between constant pressure solidification experiments and the crystallization of the inner core, as seen from the criterion of supercooling.
In contrast with usual solidification, temperature gradients are negative and destabilizing, as $\alpha$ is positive. As $m_{\mathrm{P}}G_{\mathrm{P}}$ is negative, the pressure field stabilizes the interface against topography perturbations. 

\begin{figure}
\begin{center}
\includegraphics[width = \columnwidth]{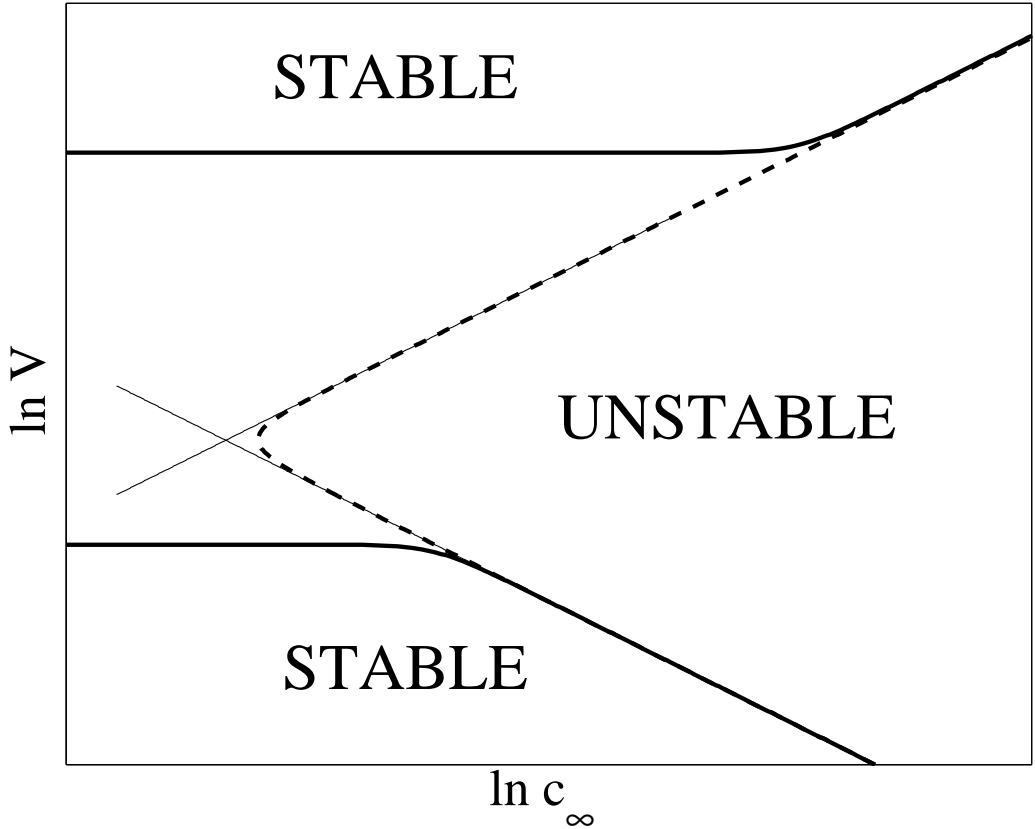}
\end{center}
\caption{Neutral curves for linear morphological stability, constructed from equation \eqref{eq:dispersion}. The dashed line is the neutral curve for the purely solutal stability problem. The solid line is the neutral curve for the thermo-solutal stability problem. \label{fig:Peffects}
}
\end{figure}

The differences between usual solidification and crystallization in a pressure field are illustrated in a $\ln V$ versus $\ln c$ diagram (figure \ref{fig:Peffects}) constructed from equation \eqref{eq:dispersion}.
The dashed curve is the neutral curve for the purely solutal stability problem: temperature gradients are taken to be constants and rejection of latent heat at the interface is neglected. 
In this limiting case, the neutral curve is very similar to that of constant pressure solidification \citep[e.g.][]{Davis01}: the curve possesses two asymptotic straight lines of slopes -1 and +1 which correspond  to the constitutional supercooling limit and to the absolute stability limit respectively, where the stabilizing effect of the surface tension becomes dominant.
In constant pressure solidification, allowing perturbations of the thermal field slightly stabilizes the interface for small $c$, and flattens the nose of the marginal curve \citep{Davis01}. 
In the system considered here, the thermal gradient is destabilizing, and even when solidifying a pure melt (no solutal destabilization), the solidification front may become unstable if the solidification velocity is high enough. 
The neutral curve of the complete thermo-solutal problem is shown as a solid line in figure \ref{fig:Peffects}.
For low $c_\infty$, the instability is mainly thermally driven and the critical solidification velocity is independent of the concentration.
As for the purely solutal destabilization, the system reaches an absolute stability limit for very high solidification rates, with the neutral curve being independent of $c_\infty$ at small concentration.

\begin{figure}
\begin{center}
\includegraphics[width = \columnwidth]{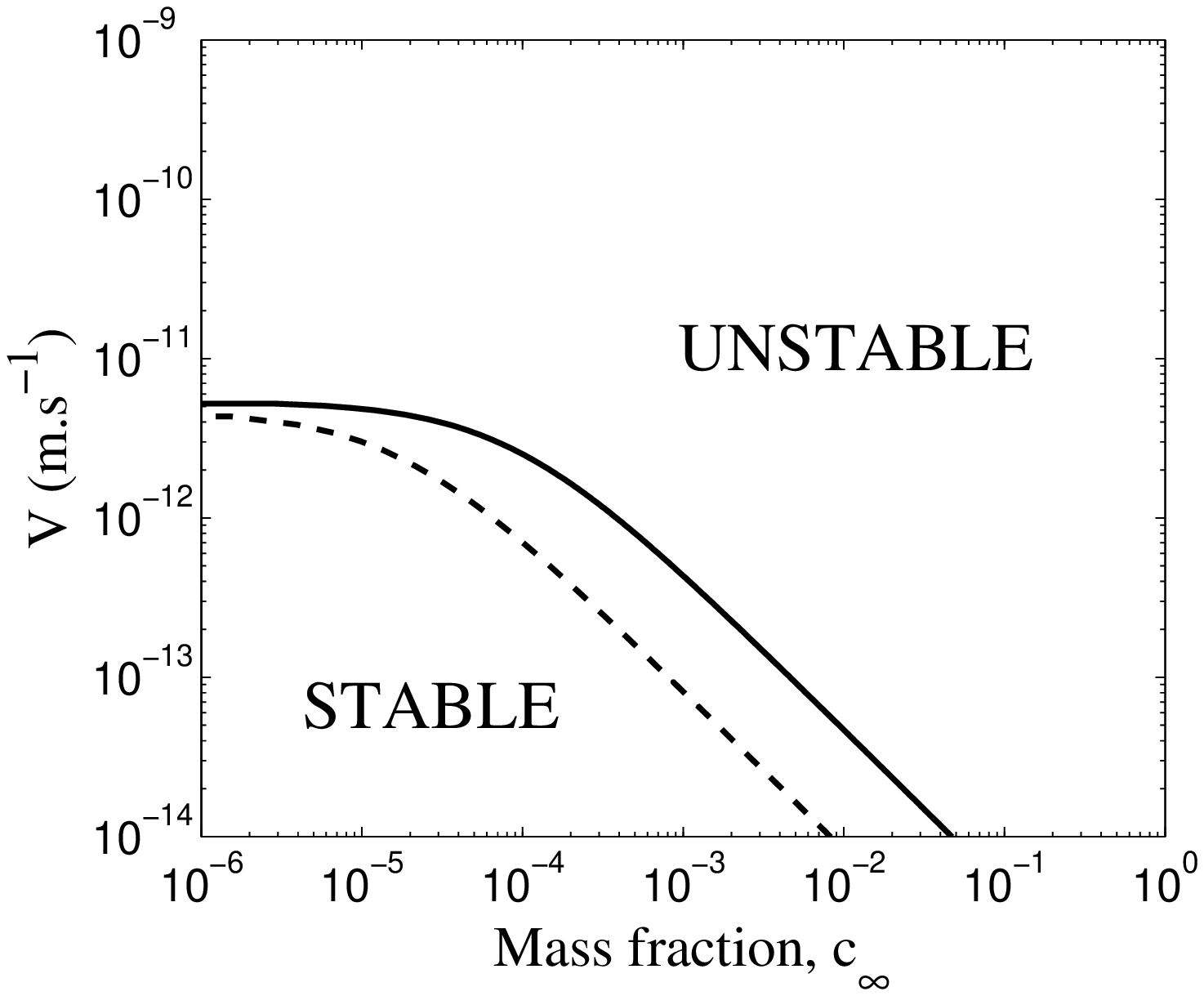}
\end{center}
\caption{Neutral curves for purely diffusive transport of solute (dashed line) and for advective transport (solid line). $m_{\mathrm c} = -10^3\ \mathrm{K}$, $k=0.2$. 
\label{fig:conv}
}
\end{figure}

Effects of advection on the stability limit is illustrated in figure \ref{fig:conv}, which represent neutral curves obtained with two different expressions for the chemical gradient at the interface. The dashed curve has been obtained with the upper bound of $G_{\mathrm c}$, in inequality \eqref{eq:bounds}, and therefore corresponds to a purely diffusive solutal transport. The solid curve has been obtained with the lower bound of $G_{\mathrm c}$, which corresponds to the case of maximum advective transport of solute. If the segregation coefficient $k$ is small, convection has a considerable stabilizing effect: at a given solute concentration $c$, the critical solidification velocity can be as much as an order of magnitude greater than in the case of no convection.

\subsection{Damping of instabilities by solid deformation}

The only effect of the pressure gradient which has been considered so far is the pressure dependence of the melting temperature.
Yet, because of the difference of density $\Delta \rho$ between the solid and the liquid, a surface topography may induced horizontal pressure gradients which may tend to flatten the interface, therefore acting against its destabilization.
To quantify the possible importance of solid flow on the interface stability, we will estimate the timescale of isostatic adjustment, assuming that the viscous deformation is driven by a balance between the pressure gradient and the viscous force:
\begin{equation}
\boldsymbol{0} = -\boldsymbol{\nabla} p + \eta \nabla^2 \boldsymbol{u},
\label{balance}
\end{equation}
where $\eta$ is the solid state viscosity.
Let us assume that the interface has a topography of amplitude $h$ which varies on a length scale $\lambda$. We are still dealing with linear stability analysis and infinitesimal perturbations, so that, as $\lambda$ is finite, 
$h \ll \lambda$. The finite amplitude case, i.e. dendrites compaction, will be considered in section 6.1.2.  
Because $h \ll \lambda$, the deformation induced by the topography must be accommodated in depth, on a length-scale $\sim \lambda$, and the horizontal and vertical velocity $u$ and $w$ must be of the same order of magnitude.
The horizontal pressure gradient is of order $\Delta \rho g h/\lambda$, $\nabla^2 u$ is of order $u/\lambda^2$, so that:
\begin{equation}
u \sim w \sim \frac{\Delta \rho g}{\eta} h \lambda.
\end{equation}
The timescale of isostatic adjustment can be defined as the ratio of the topography to the vertical velocity $\tau = h/w$, and is equal to:
\begin{equation}
\tau \sim \frac{\eta}{\Delta \rho g \lambda}.
\label{eq:isot}
\end{equation}
In the limit of $h \ll \lambda$ which is considered here, $\tau$ has a finite value even for an infinitesimal amplitude $h$ of the topography. This process is therefore relevant when dealing with linear stability analysis. 
We will however consider that this effect is negligible compared to the destabilizing effect of the thermal and compositional gradient, an assumption which will be justified a posteriori in section \ref{sect:stabICB} where $\tau$ will be compared to the timescale $1/\omega$ of the growth of an instability.

\section{Morphological stability of the ICB \label{sect5}}

\subsection{Thermo-physical parameters, composition, growth rate}

As discussed in section 2, the degree of supercooling and the stability of the boundary depend strongly on the abundance of solute and on the phase diagram of the crystallizing alloy, i.e., the liquidus slope and the segregation coefficient. The nature and relative abundances of light elements in the core are still uncertain \citep{Poirier1994}, but recent studies seem to show a preference for O, S and Si as major light elements \citep{Ringwood1991,Stixrude1997,Alfe2002,Rubie2004}.
Very little is known about the phase diagrams of the candidate alloys and it is not even clear whether those systems have an eutectic or solid-solution behavior \citep{Williams1990,Knittle1991,Boehler1993,Sherman1995,Boehler1996}.
However, a fundamental constraint on the phase diagram of the iron-major light elements system is that the outer core is richer in light solute than the inner core. 
Major light components must therefore have a segregation coefficient $k$ smaller than one and this necessarily imposes that their liquidus slopes $m_{\,\mathrm{c}}$ are negative near the Fe-rich end.

The liquidus slope and solute concentration of interest here are that of the fractionating elements. Elements which do not fractionate during the process of solidification, such as Ni and perhaps S and Si \citep{Alfe2002}, will not create chemical heterogeneity and hence will not contribute to supercooling and radial variations in melting temperature.

Assuming ideal mixing, a crude estimate of the liquidus slope of the core mixture at ICB pressure and temperature may be provided by the van't Hoff relation \citep{Chalmers1964}. We consider here the effect  of alloying Fe with a single light element, of mole fraction $x$, on the melting temperature of the mixture.
Chemical equilibrium between two multicomponents phases requires equality of the chemical potentials of each component in the two phases. In particular, the chemical potentials of the solvent, here Fe, must be equal in the liquid and in the solid: $\mu_{\mathrm {Fe}}^{\ell}=\mu_{\mathrm{Fe}}^{\mathrm s}$. In an ideal solid or liquid solution, the chemical potential of a component $i$ is expressed as $\mu_i = \mu_i^{\circ} + R T \ln x_i$, where $x_i$ is the mole fraction of component $i$ and $\mu_i^{\circ}$ is the chemical potential of pure $i$. Equality of the chemical potentials of iron in the liquid and solid phases then requires that
\begin{equation}
\mu_{\mathrm{Fe}}^{\circ \, \mathrm s} + R T \ln x_{\mathrm{Fe}}^{\mathrm s} = \mu_{\mathrm{Fe}}^{\circ \, \ell} + R T \ln x_{\mathrm {Fe}}^{\mathrm \ell},
\end{equation}
which can be rewritten, using the fact that the mole fraction of the solute is $x=1-x_{\mathrm Fe}$ and that $x^s = k x^{\ell}$, as
\begin{equation}
\frac{\mu_{\mathrm{Fe}}^{\circ \, \mathrm s}-\mu_{\mathrm{Fe}}^{\circ \, \ell}}{T} = R \ln\frac{1-x^{\ell}}{1-k x^{\ell}}.
\label{eq23}
\end{equation}
Taking the derivative of equation \eqref{eq23} with respect to $T$, and using the Gibbs-Helmoltz relation then gives
\begin{equation}
\begin{split}
\frac{\partial}{\partial T}\left(\ln\frac{1-x^{\ell}}{1-k x^{\ell}} \right)_{\mathrm P} &=\frac{\partial}{\partial T}\left( \frac{\mu_{\mathrm{Fe}}^{\circ \, \mathrm s}-\mu_{\mathrm{Fe}}^{\circ \, \ell}}{R T} \right) ,\\
&= \frac{h_{\mathrm{Fe}}^{\circ \, \mathrm s}-h_{\mathrm{Fe}}^{\circ \, \ell}}{R T_m^{\,2}} ,\\
&= \frac{M_{\mathrm {Fe}}L}{R T_m^{\,2}},
\label{eq:gibbs1}
\end{split}
\end{equation}
where $h_{\mathrm{Fe}}^{\circ \, \mathrm s}$ and $h_{\mathrm{Fe}}^{\circ \, \ell}$ are the molar enthalpy of solid and liquid Fe respectively, $L=(h_{\mathrm{Fe}}^{\circ \, \mathrm s}-h_{\mathrm{Fe}}^{\circ \, \ell})/M_{\mathrm Fe}$ is the latent heat of pure iron and $M_{\mathrm Fe}$ is the atomic weight of iron.
Assuming $k$ to be constant, for a dilute solution ($x^{\ell} \ll 1$), we obtain
\begin{equation}
\begin{split}
 \frac{\partial}{\partial T}\left(\ln\frac{1-x^{\ell}}{1-k x^{\ell}} \right)_{\mathrm P} &=\left(\frac{k}{1-kx^{\ell}}-\frac{1}{1-x^{\ell}} \right) \left.\frac{\partial x^{\ell}}{\partial T}\right|_P \\
 &\sim (k-1)\left.\frac{\partial x^{\ell}}{\partial T}\right|_P
\end{split}
\end{equation}
which together with \eqref{eq:gibbs1} yields, as $m_{\,\mathrm{c}}$ can be written as $\partial T_m/\partial x^{\ell}$,
\begin{equation}
m_{\,\mathrm{c}} \simeq \frac{R T_{\mathrm{m}}^{\circ \,2} }{M_{\mathrm{Fe}} L}(k-1) \simeq (6 \pm 3 )(k-1)\times 10^3  \mathrm{K}
\label{vanthoff}
\end{equation}
where $T_m^{\circ}$ is the melting temperature of pure iron at ICB pressure and $m_{\,\mathrm{c}}$ is given in Kelvin per atomic fraction; values and incertitudes of $T_m^{\circ}$ and $L$ are from table \ref{parameters}.
This relation means that for an ideal solution, the liquidus slope and the segregation coefficient are related by parameters which are independent of the nature of the alloying element. 
For a non-partitioning element ($k = 1$), the ideal liquidus slope is equal to zero, whereas a highly partitioning element ($k$ small compared to 1) will have a high liquidus slope.
Through this relation, constraints on $k$ may provides informations on the liquidus slope of the core mixture. 

From estimates of the volume change during melting and of the density jump at the inner core boundary, \citet{Anderson1994} estimated the ratio of light elements in the outer core to that in the inner core to be approximately 4 to 1.
This gives a global segregation factor $k=0.25$ and, by equation \eqref{vanthoff}, a liquidus slope $m_c \simeq -4.5 \pm 2.5 \times 10^3\ \mathrm{K} $. 
Note that the segregation factor estimated here is an effective segregation factor, which is higher than the thermodynamic one, and should give a lower bound of $|m_c|$.
This estimate may be appropriate if there is only one light element in the core, but finer estimates are needed if there are several light elements of comparable concentrations.
As an example, the {\it ab initio} simulations of \citet{Alfe2002} suggest that the outer core may be composed of $\simeq 10$ mole\% of S and/or Si and $\sim 8$ mole\% ($\simeq 2$ wt.\%) of O.
According to \citet{Alfe2002}, S and Si do not significantly fractionate and the density jump at the ICB may be accounted for by fractionation of oxygen alone, whose segregation coefficient has been estimated to be 0.02. 
With this value, $m_c$ tends to its $k=0$ bound $m_c \simeq -6 \pm 3 \times 10^3\ \mathrm{K} $. 
In what follows, two chemical models of the core will be considered: one with a single light element of concentration $\simeq 10$ wt.\%, and the other with only one fractionating light element (but several light elements), oxygen, of concentration $\simeq 2$ wt.\%.

Our relation is similar to the one derived by \citet{Alfe2002}. \citet{Stevenson1981} and \citet{Anderson1997} estimated the melting point depression by assuming equilibrium between a pure solid and an alloyed liquid, and therefore obtained the upper bound of our estimate (i.e. our $k=0$ value). 
Those theoretical estimates are in poor agreement with experiments.
Experimental results on the melting temperature of the Fe-O system \citep{Knittle1991,Boehler1993} predict a small melting temperature depression, and perhaps a solid-solution behavior.
The case of the Fe-S system if more controversial. 
\citet{Williams1990} results suggest that the eutectic behavior persists at high pressure, and predict a significant melting point depression ($m_{\mathrm c}$ is of order $-5\times10^3$ K at core-mantle boundary pressure).
In contrast \citet{Boehler1996} found that the Fe-FeS eutectic melting depression becomes much smaller at high pressures, and conclude that this supports the possibility of solid-solution between Fe and FeS at core pressures.
To our knowledge, there is no experimental work at this pressure range dealing with other candidate alloys.
In the present work, values of $m_{\mathrm c}$ between $-10^2$ K and $-10^4$ K will be considered.

The interfacial energy of iron at ICB conditions can be deduced from estimates of the latent heat of crystallization, because those two parameters both derive from the difference of atoms bonds energy between the solid and liquid phase.
The interfacial energy per atom $\gamma_a$ can be calculated to be 1/4 of the atomic latent heat $L_a$ for a flat close-packed surface \citep{Chalmers1964}.
Estimates of the latent heat of iron at ICB conditions range from $600\, \mathrm{ kJ\,kg^{-1}}$ to  $1200\, \mathrm{ kJ\,kg^{-1}}$ \citep{Poirier1994a,Anderson1997,Laio2000,Vocadlo2003} and
from these values, we estimate the interfacial energy per unit area $\gamma$ to be $0.4\pm0.2\ \mathrm{J\, m^{-2}}$, which can be compared to the $0.204$ J.m$^{-2}$ value at standard conditions \citep{Chalmers1964}.

\citet{Buffett1992} proposed an analytical model of growth of the inner core and found that the radius increase at leading order as the root square of time, $r = r_{\mathrm {ic}} \sqrt{t/a}$, where $r_{\mathrm {ic}}$ is the present radius of the inner core and $a$ is its age. The present solidification velocity is then $V = r_{\mathrm {ic}}/2a$. 
If the inner core is young \citep[e.g.][]{Labrosse2001,Nimmo2004}, i.e. $a \sim 1$ Ga, V is found to be $2\times10^{-11}$ m.s$^{-1}$. On the other hand, if the inner core nucleated around 3 Ga ago, as \citet{Christensen2004} claim, V could be of order $6\times10^{-12}$ m.s$^{-1}$.
\citet{Wen06} observed a temporal change of travel time of the PKiKP phase between the two events of an earthquake doublet, indicating a localized change of the inner core radius of about 1 km in ten years.
This observation may be interpreted as reflecting episodic growth of the inner core, coupled with non stationary convection in the outer core \citep{Wen06}.
The resulting instantaneous solidification velocity is $\simeq 10$ km/10 years $\simeq 10^{-6}$ m.s$^{-1}$, which is much higher than the mean solidification velocity estimated from models of the core thermal history.

Other parameters used in this study, with values currently found in the literature, are listed in table \ref{parameters}.

\begin{table*}[ht] 
\caption{Parameters used in this study. \label{parameters}}
\begin{center}
\begin{tabular*}{\textwidth}{@{\extracolsep{\fill}}lll}
\hline 
Clapeyron slope & $m_{\mathrm P}$ & $\simeq 10^{-8}\ \mathrm{K\, Pa^{-1}}$ $^a$\\
Liquidus slope & $m_{\mathrm c}$ &  $-10^{2}$ to $-10^{4}\ \mathrm{K}$ $^b$\\
Light elements concentration & $c_\infty$ & 2 to 10 wt.\% $^b$\\
Growth rate of the inner core & $V$ & {$6 \times 10^{-12}$} to $2\times10^{-11}\ \mathrm{m\, s^{-1}}$ $^b$\\ 
Temperature at the ICB & $T_{\mathrm{icb}}$ & $5000$ to $6000\ \mathrm{K}$ $^{a, c}$\\
Specific heat at constant pressure & $c_P$ & $860\ \mathrm{J \, kg^{-1}\, K^{-1}}$\\
Latent heat of crystallization & $L$ & $600$ $^c$ to $1200\ \mathrm{kJ\, kg^{-1}}$ $^a$\\
Thermal conductivity in the liquid & $\kappa^\ell$ & $63\ \mathrm{W\, m^{-1}\, K^{-1}}$ $^d$\\
Thermal conductivity in the solid & $\kappa^s$ & $79\ \mathrm{W\, m^{-1}\, K^{-1}}$ $^d$\\
Thermal diffusivity in the liquid & $D_{\mathrm T}^\ell$ & $6\times10^{-6}\ \mathrm{m^2\, s^{-1}}$ $^d$\\
Thermal diffusivity in the solid & $D_{\mathrm T}^s$ & $7\times10^{-6}\ \mathrm{m^2\, s^{-1}}$ $^d$\\
Chemical diffusivity & $D_c$ & $10^{-9}\ \mathrm{m^2\, s^{-1}}$ $^c$\\
Viscosity of the inner core & $\eta$ & $10^{16}$ to $10^{21}\ \mathrm{Pa\, s}$ $^b$\\
Liquid-solid interfacial energy & $\gamma$ & $0.4\pm0.2\ \mathrm{J\, m^{-2}}$ $^b$\\
Gibbs-Thomson coefficient & $\Gamma$ & $\simeq 2\times 10^{-7}$ K.m $^b$\\
\hline
\end{tabular*}
\end{center}
$^a$ from \citet{Anderson1997}. \\
$^b$ see text. \\
$^c$ from \citet{Poirier1994a}. \\
$^d$ from \citet{Stacey}. \\
\end{table*}

\subsection{Supercooling and stability analysis  \label{sect:stabICB} }

The liquid at the ICB is supercooled if condition \eqref{eq:crit} is satisfied. This criterion can be rewritten, using the heat balance at the interface (equation \eqref{eq:heat_balance}), as:
\begin{equation}
\frac{L_v V}{\kappa^\ell} + m_{\,\mathrm{c}}\, G_{\mathrm c} -\frac{\kappa^s}{\kappa^\ell} G_{\mathrm T}^s > \rho g m_{\,\mathrm{P}}.
\end{equation}
Depending on the age of the inner core, on the value of the thermal conductivity and on the hypothetical presence of radioactive elements, the thermal gradient $G_{\mathrm T}^{\mathrm s}$ in the inner core at the ICB may vary widely \citep{Yukutake1998}. 
However, the thermal term $L_v V / \kappa^\ell$, which is the contribution to supercooling from the heat released by crystallization, is high enough to balance the pressure term if the solidification velocity is higher than $\sim 5 \times 10^{-12}$ m.s$^{-1}$.
This is roughly equal to the lowest estimates of $V$, and, as the term $-G_{\mathrm T}^s$ is positive, it is thus likely that thermal terms alone are sufficiently high to ensure supercooling.
The chemical term $m_{\,\mathrm{c}}\, G_{\mathrm c}$ is probably much higher, at least one order of magnitude greater than the pressure term if $m_{\mathrm c}$ is as small as $-10^2$ K and three order of magnitude greater if $m_{\mathrm c}=-10^4$ K, and the conclusion that the vicinity of the ICB is indeed supercooled seems to be inescapable.

\begin{figure}
\begin{center}
\includegraphics[width = \columnwidth]{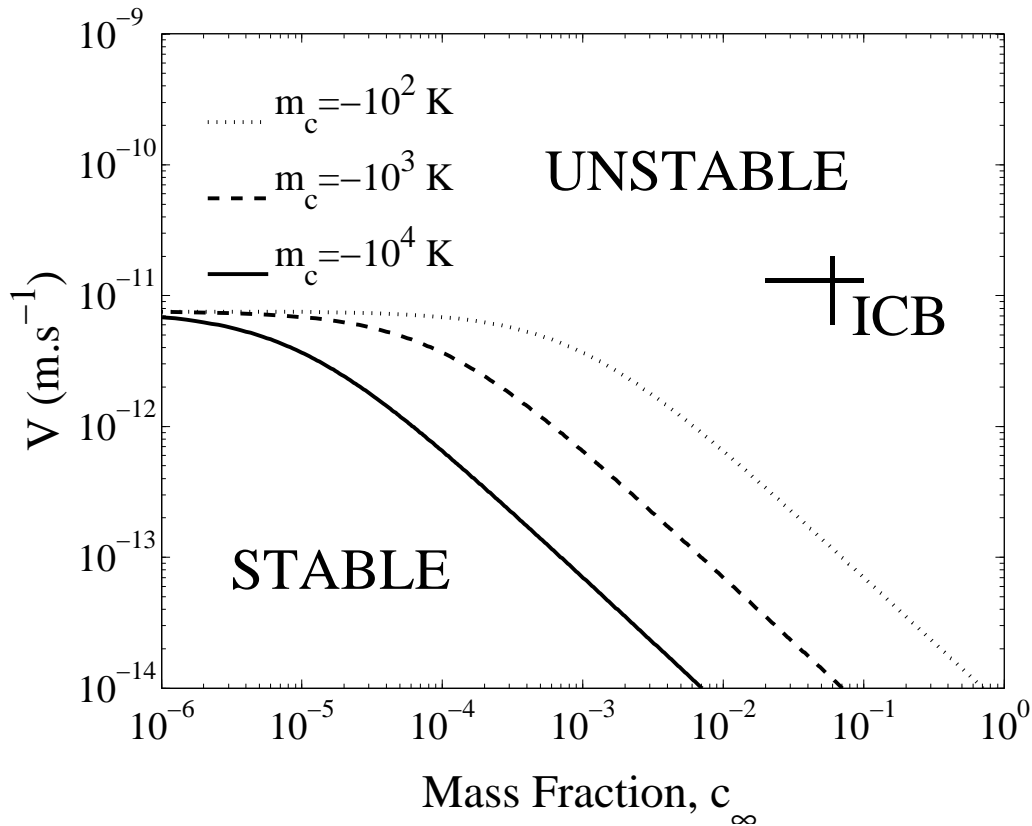}
\end{center}
\caption{Neutral curves for linear morphological stability, with $m_{\mathrm c}=-10^2\ \mathrm{K}$ (dotted line), $m_{\mathrm c}=-10^3\ \mathrm{K}$ (dashed line) and $m_{\mathrm c}=-10^4\ \mathrm{K}$ (solid line). $G_{\mathrm T}^{\mathrm s}=0$ K.m$^{-1}$ and $k=0.25$. The straight lines cross is the location of the ICB in the stability diagram. Uncertainties in $V$ result from uncertainties in the age of the inner core. The value of $c_\infty$ depends on the chemical model chosen: $c_\infty\simeq10$ wt.\% if there is only one dominant light element, and $c_\infty\simeq2$ wt.\% if the model of \citet{Alfe2002} is adopted.
\label{stabICB}
}
\end{figure}

We now use the dispersion equation \eqref{eq:dispersion} to investigate more quantitatively the stability of an initially plane solid-liquid interface in the actual conditions of the ICB.
As discussed in section 3, we choose to take the lower bound of the chemical gradient, given by equation \eqref{eq:lowerbound}, so that the effect of convection on the basic state may be overestimated rather than underestimated.
We tested several values of each parameter within their uncertainty ranges and found that for all plausible sets of values, the ICB is unstable. 
The liquidus slope is by far the most critical parameter. Because the chemical gradient is proportional to $m_{\mathrm c}$, the two orders of magnitude uncertainty on $m_{\mathrm c}$ propagate directly in the uncertainties on the location of the marginal stability curve.
Neutral curves for linear stability are plotted in figure \ref{stabICB}, with $m_{\mathrm c}=-10^2$ K, $-10^3$ K and $-10^4$ K, $G_{\mathrm T}^{\mathrm s}=0$ K.m$^{-1}$ and $k=0.25$. 
Despite the strong dependence on $m_{\mathrm c}$ of the neutral curve location, the solidification velocity is two orders of magnitude greater than the critical velocity for a liquidus slope as small as $-10^2$ K, and three or four orders of magnitude greater than the critical velocity for a liquidus slope of $-10^3$ K or $-10^4$ K.
If the growth of the inner core is episodic, with a much higher instantaneous solidification velocity, the ICB would be even more unstable.

\begin{figure}
\begin{center}
\includegraphics[width = \columnwidth]{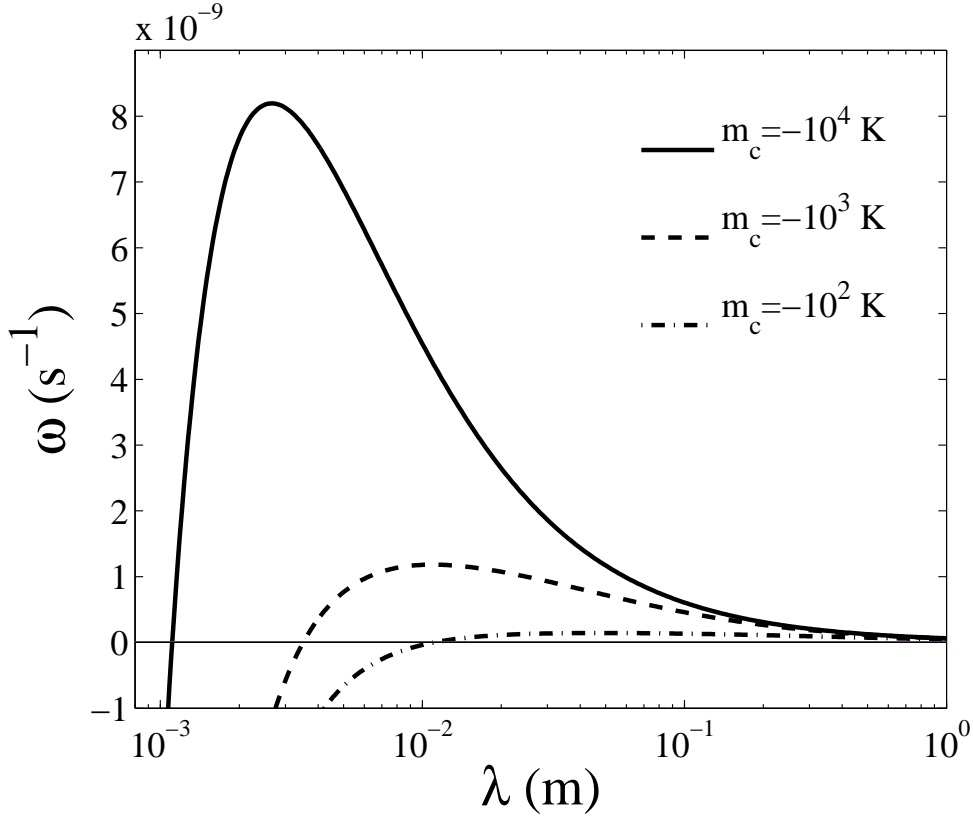}
\end{center}
\caption{ Growth rate of infinitesimal perturbations against wave length, at the conditions of the ICB, for $m_{\mathrm c} = -10^4$ K (solid line), 
$-10^3$ K (dashed line) and $-10^2$ K (dash-dotted line). $k=0.02$ and $L_v=600$ kJ.kg$^{-1}$.
}
\label{growth_rates}
\end{figure}

In figure \ref{growth_rates}, the growth rate of infinitesimal perturbations has been plotted against wave length, at the conditions of the ICB, for three different values of $m_c$.
The corresponding time-scales range from a few years to about 300 years.
We may now estimate the timescale of isostatic adjustment, given by equation \eqref{eq:isot}, and compare it to the timescale of instability growth.
The solid inner core viscosity is poorly constrained: estimates range from $10^{16}$ Pa.s to more than $10^{21}$ Pa.s \citep{Buffett1997, Yoshida1996}. A lower estimate of the viscous flow timescale may be given by taken $\eta=10^{16}$ Pa.s. With $\Delta \rho = 600$ kg.m$^{-3}$, $g=4.4$ m.s$^{-1}$ \citep{PREM}, the isostatic adjustment timescale is of order $10^{14}$ s $\simeq 3\times 10^6$ years  for $\lambda \sim 1$ cm, about four or six orders of magnitude greater than the timescale of instability growth: isostatic adjustment will not delay the instability.
Our results are consistent with the conclusion of \citet{Shimizu2005} that the timescale of dendrites growth is very short compared to the timescale of inner core growth, suggesting that a mushy layer will indeed form at the ICB.


Results have been presented here for parameters (growth rate, thermal and pressure gradient) inferred for the present state of the inner core, 
but when dealing with the internal structure of the inner core, it is of equal interest to investigate what was the solidification regime during its past history.
As explained before, the solidification regime depends primarily upon the thermal, solutal and pressure gradients at the interface, the thermal and solutal fields being destabilizing whereas the pressure field is stabilizing.
Because the solidification velocity was most certainly greater in the past than it is today, the rates of release of heat and solute were also greater; thermal and solutal gradients were therefore steeper (more destabilizing).
In addition, the liquidus slope $m_{\,\mathrm{P}} G_{\,\mathrm{P}}$ is less steep at deeper depth (because the gravity field, and hence the pressure gradients decrease to zero at the center of the Earth), and therefore less stabilizing.
At first order, all terms seem to act in the same way, and
it is then likely that the solidification of the inner core has been dendritic for most of its history.

\section{Length scales of the mush}

\subsection{Vertical length scales}

\subsubsection{Thermodynamic depth of the mushy zone}

In the laboratory, and in metallurgical applications, the mushy zone depth is typically a few centimeters.
However, because of the very small temperature gradient in the inner core, and of the additional effect of pressure, the temperature in the inner core remains close to the melting temperature, which suggests that liquid enriched in solute may remain thermodynamically stable at considerable depths \citep{Fearn1981}. 

While freezing occurs in the mushy zone, interdendritic melt is further enriched in solute, lowering its melting temperature further.
An enriched liquid phase can coexist with the solid phase as long as the actual temperature is above the melting temperature. The depth $\delta_{\mathrm{Th}}$ at which a liquid of concentration $c = c_o + \Delta c$ can be in thermodynamic equilibrium with the surrounding solid phase is then given by equating the actual and melting temperatures:
\begin{equation}
T(\delta_{\mathrm{Th}})=T_m(\delta_{\mathrm{Th}}).
\end{equation}
To a good approximation, the acceleration of gravity is linear in $r$ in the inner core, so that pressure is quadratic in radius.
Assuming for convenience the temperature profile to be quadratic as well (which is a not so bad approximation if the cooling rate is approximately constant within the inner core), the following expression for $\delta_{\mathrm{Th}}$ can be found:
\begin{equation}
\frac{\delta_{\mathrm{Th}}}{r_{\mathrm{ic}}} \simeq \sqrt{\frac{ |m_c| \Delta c}{\Delta \Theta}},
\label{eq:d_mush}
\end{equation}
where $\Delta \Theta\simeq 150\ \hbox{to}\ 200$ K is the difference between the actual temperature and the melting temperatures (at outer core composition) at the center of the inner core \citep{Yukutake1998}.
The precise form of the temperature profile (quadratic or not) is not of great importance for our order of magnitude estimates.

The mushy zone can extend to the center of the inner core ($\delta_{\mathrm{Th}}/r_{\mathrm{ic}}=1$) if the variation of melting temperature with pressure is compensated by $|m_{\mathrm c}| \Delta c$.
The maximum allowable light elements concentration is the concentration at the eutectic (if it exists), and the maximum value of $|m_{\mathrm c}| \Delta c$ is therefore $|m_{\mathrm c}|(c_{\mathrm E}-c_0)$. Taking $c_{\mathrm E}=25$ at.\% as a plausible value \citep{Stevenson1981}, we found from equation \eqref{eq:d_mush} that $\delta_{\mathrm{Th}}=r_{ic}$ if $|m_{\mathrm c}|$ is greater than $1.3 \times 10^3$ K. On the other hand, if $|m_{\mathrm c}|$ is as small as $10^2$ K, the resulting depth of liquid thermodynamic equilibrium is of the order of 300 km.

\subsubsection{A collapsing mushy zone?}

As noted by \citet{Loper1983}, a fundamental observation is that the ICB appears to be sharp on seismic wavelength scale ($\sim 10$ km), which means that the solid fraction at the top of the inner core must become significant within a few kilometers in depth, a fact that seems at first view hard to reconcile with the presence of a hundred kilometers thick mushy zone.
\citet{Loper1983} argued that convective motions in the mush may cause the dendrites to thicken and calculated that the solid fraction may become of order one at $\simeq 300$ m below the top of the mushy zone, thus explaining the seismic observations.

In addition to this process, the solid mass fraction may also increase in depth because of gradual collapse of the dendrites under their own weight.
The physical process at stake is quite similar to the isostatic adjustment considered in section \ref{section4}: the density difference between the solid and the liquid implies horizontal pressure gradients which increase with depth in the mushy zone, making the dendrites to broaden at their base by viscous solid flow.
The timescale of dendrites widening may be estimated by considering equilibrium between the horizontal pressure gradient and the viscous force in the solid (equation \eqref{balance}), as was done is section \ref{section4}.
Here the horizontal length scale $\lambda$, say the diameter of a dendrite, is supposed to be small compared to the height $h$ of a dendrite. As before, the horizontal velocity at depth $h$ is of order:
\begin{equation}
u \sim \frac{\Delta \rho g}{\eta} h \lambda,
\end{equation}
but the timescale of interest is now the ratio of horizontal length scale to the horizontal velocity $\tau_{c} = \lambda/u$, so that the timescale of dendrite widening - or compaction - is:
\begin{equation}
\tau_{c} \sim \frac{\eta}{\Delta \rho g h},
\end{equation}
which is inversely proportional to the height of the dendrite.
As long as the timescale of dendrites growth is small compared to $\tau_{\mathrm c}$, the depth of the mushy zone increases.
When increasing the height of the dendrites, the timescale of dendrites widening decreases, and a stationary state is eventually reached, where the growth of the dendrites by solidification at their tips is balanced by the collapse of the mushy zone.
This stationary depth is then found by equating $\tau_{\mathrm c}$ to the timescale of dendrites growth, which is $h/V$ when the system is stationary. This yields:
\begin{equation}
h \sim \sqrt{\frac{\eta V}{\Delta \rho g}}.
\end{equation}
Uncertainties in $h$ come mostly from uncertainties in the viscosity. With $\eta=10^{16}$ Pa.s, $h$ is found to be a few meters, a very small value, whereas $\eta=10^{21}$ Pa.s leads to $h\sim 1$ km \citep{Sumita96}.
Although estimates of $h$ span more than two orders of magnitude, they are in any case significantly smaller than the thermodynamic estimate of the depth of the mushy zone.
This means that as a result of compaction, the solid fraction may become of $O(1)$ at a few $h$ in depth, well before the thermodynamic limit of liquid equilibrium is reached. 

\subsection{Horizontal length scale: interdendritic spacing}

Interdendritic spacing $\lambda_1$ (the distance between two dendrites tips) is a length scale of considerable interest for the structure of the inner core.
Because each columnar crystal is usually made of many dendrites \citep{Kurz89}, the interdendritic spacing gives a lower bound of the grain size, and hence may gives hints on the deformation mechanism and viscosity relevant to the possible viscous deformation of the inner core.
In addition, the primary dendrite spacing is needed for estimating the permeability of the mush, which is roughly proportional to $\lambda_1^2$ \citep{Bergman94}. Permeability is a necessary parameter for the study of the mush hydrodynamics \citep[e.g.][]{Bergman94,LeBars06}, or for quantitative compaction models \citep{Sumita96}.

We make the assumption that dendrites have an axisymmetric shape described by a function $g(z)$, which is linked to the solid fraction $f(z)$ by the relation:
\begin{equation}
 f(z)=\frac{\pi g^2(z)}{A(\lambda_1)},
 \label{solid_fraction}
\end{equation}
where $A(\lambda_1)$ is the horizontal surface area occupied by a dendrite. In a cubic dendritic array, $A(\lambda_1)$ is simply $\lambda_1^2$, whereas in a hexagonal dendritic array, $A(\lambda_1)=3/2\tan(\pi/3)\lambda_1^2\simeq 0.86 \lambda_1^2$.

$f$ and $g^2$ are proportional through a constant involving $\lambda_1$, so that if $f$ is known, an additional constraint on $g$ is enough to determine $\lambda_1$.
This constraint can be given by Langer and M{\"u}ller-Krumbhaar's theory of dendrite tip radius selection \citep{Langer77}, in which the dendrite tip radius $R$ is equal to the shortest wavelength $\lambda_i$ for which the interface is unstable.
This theory is in very good agreement with experiments, and we will follow it in the present work.
It can be shown \citep[see][]{Kurz89} that at low solutal Peclet number (Pe$_c=V R/D \ll 1$), a condition well satisfied here, the chemical gradient $G_c$ at the dendrite tip is equal to the lower bound of the chemical gradient we  used in our stability analysis (equation \eqref{eq:lowerbound}). Estimates of $\lambda_{i}$ from our stability analysis thus directly give the adequate $R$.
A good approximation for $R$ at low solutal and thermal Peclets numbers can be derived from equation \eqref{eq:dispersion}, noting that the chemical term in the numerator of \eqref{eq:dispersion} is much bigger than the thermal and pressure terms \citep{Kurz89}:
\begin{equation}
  R \simeq 2\pi \sqrt{\dfrac{\Gamma D}{m_{\mathrm{c}} c_{\infty} (k-1) V}} 
\end{equation}
$R$ ranges from $\sim 1$ mm if $m_{\mathrm c}=-10^4$ K to $\sim 1$ cm if $m_{\mathrm c}$ is as low as $-10^2$ K (see figure \ref{growth_rates}). Uncertainties from other parameters than $m_{\mathrm c}$ are much smaller.

On the other hand, the dendrite tip radius $R$ is equal to the radius of curvature of $g$ taken in $z=0$ which is by definition equal to:
\begin{equation}
 R=-\frac{(1+g'(0)^2)^{3/2}}{g''(0)}
 \label{curvature_radius}
\end{equation}
where $R$ is defined to be positive.
Inserting equation \eqref{solid_fraction} into \eqref{curvature_radius}, with the boundary condition $f(0)=0$, yields
\begin{equation}
 A(\lambda_1) = \frac{2 \pi R}{\left.\dfrac{\partial f}{\partial z}\right|_\mathrm{ICB}}.
 \end{equation}
 or, for an hexagonal dendritic array:
\begin{equation}
 \lambda_1 = \sqrt{\dfrac{2 \pi R}{0.86\left.\dfrac{\partial f}{\partial z}\right|_\mathrm{ICB}}} .
 \label{lambda}
\end{equation}
Although simple, this relation is very general, the only assumptions made here being that of stationary state and that of an axisymmetric dendrite tip.
If the appropriate assumptions on solute transport and selection of dendrites tip radius are made, classical models of primary dendrite spacing \citep{Trivedi84,Hunt79} follow from equation \eqref{lambda}.

These models have been derived for non-convecting mushy zones where the solid fraction increase smoothly from zero to one at the thermodynamic depth of the mush, that is, $\partial f/ \partial z |_\mathrm{ICB}$ scales as $1/\delta_{\mathrm{Th}}$ \footnote{Note that, in laboratory experiments, with no pressure gradient and with a constant temperature gradient $\delta_{\mathrm{Th}}\sim m_{\mathrm c} (c_e-c_{\infty})/G_{\mathrm T}$, so that, as $R \propto V^{-1/2}$, $\lambda_1 \propto G_{\mathrm T}^{-1/2}V^{-1/4}$. This scaling law has been shown to be in very good agreement with experiments \citep[e.g.][]{Kurz89}.}. 
However, seismology tells us that the ICB is sharp on a scale of 10 km, which means that the scale length of solid fraction increase at the top of the inner core is obviously not the thermodynamic mushy zone depth, and must be at maximum $\sim 10$ km. Therefore a minimum order of magnitude of $\partial f/ \partial z |_\mathrm{ICB}$ is $1/10$ km $=10^{-4}$ m$^{-1}$. This may give upper bounds for $\lambda_1$; we found $\lambda_1\sim 10$ m if $m_{\mathrm c}=-10^4$ K and $\lambda_1\sim 30$ m if $m_{\mathrm c}=-10^2$ K.
Tighter constraints may be found if we could estimate more precisely $\partial f/ \partial z |_\mathrm{ICB}$.
Compaction is not expected to affect significantly the dendrites spacing because it is not effective in the very top of the mush. The solid fraction profile in the upper few meters, from which the primary spacing results, is more probably controlled by convection.
In his convective mushy zone model, \citet{Loper1983} found that $\partial f/ \partial z |_\mathrm{ICB}\sim 3\times 10^{-3}$ m$^{-1}$.
With this value and our estimates of $R$, we found $\lambda_1$ to be 1 to 5 meters. In Loper's estimate, $\partial f/ \partial z |_\mathrm{ICB}$ is proportional to the typical velocity $W$ of descending liquid close to the ICB, taken to be around $10^{-6}$ m.s$^{-1}$. This value is quite uncertain, but as $\lambda_1$ is inversely proportional to the square root of $\partial f/ \partial z |_\mathrm{ICB}$, a change of two orders of magnitude in $W$ is needed in order to change the order of magnitude of $\lambda_1$.
As a consequence, and considering all sources of uncertainties, it appears difficult to have an interdendritic spacing of more than a few tens of meters, our preferred estimate being a few meters.
If the growth of the inner core is episodic, the interdendritic spacing is expected to be smaller. As $\lambda_1 \propto V^{-1/4}$, taking $V=10^{-6}$ m.s$^{-1}$ instead of $V=10^{-11}$ m.s$^{-1}$ would result in an interdendritic spacing about twenty times smaller than estimated above, that is, a few tens of centimeters.

Our estimates are significantly smaller than estimate of a few hundred meters from scaling laws \citep{Bergman1998}.
This is not surprising because those scaling laws have been derived for non-convecting (and non-compacting) mushy zones, where $f$ is linearly increasing in the whole thermodynamic mushy zone.
Assuming for a demonstrative purpose that $f$ increases linearly from 0 to 1 within the thermodynamic mushy zone, whose depth is taken equal to the inner core radius, equation \eqref{lambda} gives $\lambda_1 \sim 100 - 300$ m, in very good agreement with Bergman's estimate.

Our assumption of an axisymmetric dendritic shape is justified if the iron phase at inner core conditions is fcc or bcc iron \citep{Vocadlo2003b}, but might not hold if it is $\epsilon$-iron, as hcp materials (e.g. ice or zinc) usually have plateshaped dendrites, known as platelets \citep{Bergman2003}.
If iron dendrites are indeed platelets, estimating the inter-platelets spacing would require some modifications of our analysis to take into account the specific geometry of platelets.
Although this is expected to give quantitatively different results, this should not alter the qualitative conclusion that convection in the mush reduces the interdendritic spacing, and that interdendritic spacing at ICB should be much smaller than suggested by classic scaling laws.

\section{Conclusion}

The morphological stability of an initially plane solidification front at the ICB conditions has been investigated. Despite the stabilizing effects of convection and of the pressure gradient, a continuous solidification implies non-zero solutal and thermal gradients at the ICB which, for plausible parameters values, are high enough for the interface to be destabilized.
Because the conditions in the past were even more destabilizing, it is probable that the ICB has been dendritic through most of the inner core history.

Thermodynamic considerations predict a very thick mushy zone which, as noted by \citet{Fearn1981}, could possibly extend to the center of the Earth. However, considerable uncertainties on the phase diagram do not allow a precise estimate of the thermodynamic depth, which may be only a few tens of kilometers if the liquidus slope is small.
The most superficial part of the inner core may be understood as a collapsing mushy zone, where both convection and compaction act to rapidly increase the solid fraction within a length scale probably smaller than 1 km, making the ICB to appear seismically sharp.
The length scale of the thermodynamic depth of the mushy zone and the compaction length scale, although clearly different, are unconstrained by the current knowledge of the phase diagram of the core mixture and of the solid iron viscosity at inner core conditions. 
Progress in the determination of these parameters may greatly help the understanding of the inner core structure.

We tried to constrain the primary dendrite spacing $\lambda_1$ of the mushy zone. $\lambda_1$ appears to depend on the vertical derivative of the solid fraction at the top of the mush which, in turn, depends on the vigor of convection in the mush. Here again, it is difficult to make precise and reliable estimates, but we found that the interdendritic spacing is most probably smaller than a few tens of meters, and possibly only a few meters.

Whether or not a significant amount of melt may subsists at large depth remains an open question.
If the mush permeability is of order $\lambda_1^2/100$ \citep{Bergman94}, our estimate of $\lambda_1$ suggests permeability values higher than $10^{-2}$ m$^2$. The work of \citet{Sumita96} suggests that with such a high permeability, the liquid will be very efficiently removed from the inner core by compaction, and that the residual liquid fraction will be essentially zero.
A high permeability in the mush does not however rule out the possibility that unconnected, trapped liquid pockets persist in depth.
The answer to this question may depend in part on surface tension driven processes.
Sintering, i.e. migration of grain boundaries driven by surface tension, may play an important role in the redistribution of the liquid phase and have consequences on the efficiency of compaction, in a way which will depend on the wetting properties of the melt.
If the (unknown) dihedral angle of light elements rich liquid iron in contact with solid iron is greater than $60^\circ$ \citep{Bulau79}, the liquid phase is expected to become unconnected at a given liquid fraction, thus living a residual liquid phase in the inner core.

\begin{ack}

We thank the two anonymous reviewers for their careful reviews and constructive suggestions.
We are grateful to Thierry Duffar, Hisayochi Shimizu, Jean-Louis Le Mou{\"e}l and Jean-Paul Poirier for helpful discussions.
Careful reading and useful comments by Dominique Jault, Alexandre Fournier, Philippe Cardin, Franck Plunian and Elisabeth Canet were much appreciated.

\end{ack}

\bibliography{biblio}

\begin{thebibliography}{52}
\expandafter\ifx\csname natexlab\endcsname\relax\def\natexlab#1{#1}\fi
\expandafter\ifx\csname url\endcsname\relax
  \def\url#1{\texttt{#1}}\fi
\expandafter\ifx\csname urlprefix\endcsname\relax\def\urlprefix{URL }\fi

\bibitem[{{Alf{\`e}} et~al.(2002){Alf{\`e}}, {Gillan}, and {Price}}]{Alfe2002}
{Alf{\`e}}, D., {Gillan}, M.~J., {Price}, G.~D., 2002. {Ab initio chemical
  potentials of solid and liquid solutions and the chemistry of the Earth's
  core}. J. Chem. Phys. 116, 7127--7136.

\bibitem[{{Anderson} and {Duba}(1997)}]{Anderson1997}
{Anderson}, O.~L., {Duba}, A., 1997. {Experimental melting curve of iron
  revisited}. J. Geophys. Res. 102, 22659--22670.

\bibitem[{Anderson and Ahrens(1994)}]{Anderson1994}
Anderson, W., Ahrens, T., 1994. {An equation of state for liquid iron and
  implications for the Earth's core}. J. Geophys. Res. 99, 4273--4284.

\bibitem[{{Bergman}(1998)}]{Bergman1998}
{Bergman}, M.~I., 1998. {Estimates of the Earth's inner core grain size}.
  Geophys. Res. Lett. 25, 1593--1596.

\bibitem[{{Bergman} et~al.(2003){Bergman}, Agrawal, Carter, and
  M.}]{Bergman2003}
{Bergman}, M.~I., Agrawal, S., Carter, M., M., M.-S., 2003. {Transverse
  solidification textures in hexagonal close-packed alloys}. J. Crystal Growth
  255, 204--211.

\bibitem[{{Bergman} and {Fearn}(1994)}]{Bergman94}
{Bergman}, M.~I., {Fearn}, D.~R., 1994. {Chimneys on the Earth's inner-outer
  core boundary?} Geophys. Res. Lett. 21, 477--480.

\bibitem[{{Boehler}(1993)}]{Boehler1993}
{Boehler}, R., 1993. {Temperatures in the Earth's core from melting-point
  measurements of iron at high static pressures}. Nature 363, 534--536.

\bibitem[{{Boehler}(1996)}]{Boehler1996}
{Boehler}, R., 1996. {Fe-FeS eutectic temperatures to 620 kbar}. Phys. Earth
  Planet. Inter. 96, 181--186.

\bibitem[{Braginsky(1963)}]{Braginsky1963}
Braginsky, S., 1963. {Structure of the F layer and reasons for convection in
  the Earth's core}. Dokl. Akad. Nauk. SSSR Engl. Trans. 149, 1311--1314.

\bibitem[{{Buffett}(1997)}]{Buffett1997}
{Buffett}, B.~A., 1997. {Geodynamics estimates of the viscosity of the Earth's
  inner core}. Nature 388, 571--573.

\bibitem[{{Buffett} et~al.(1992){Buffett}, {Huppert}, {Lister}, and
  {Woods}}]{Buffett1992}
{Buffett}, B.~A., {Huppert}, H.~E., {Lister}, J.~R., {Woods}, A.~W., 1992.
  {Analytical model for solidification of the Earth's core}. Nature 356,
  329--331.

\bibitem[{{Bulau} et~al.(1979){Bulau}, {Waff}, and {Tyburczy}}]{Bulau79}
{Bulau}, J.~R., {Waff}, H.~S., {Tyburczy}, J.~A., Oct. 1979. {Mechanical and
  thermodynamic constraints on fluid distribution in partial melts}. J.
  Geophys. Res. 84, 6102--6108.

\bibitem[{{Cao} and {Romanowicz}(2004)}]{Cao2004}
{Cao}, A., {Romanowicz}, B., 2004. {Hemispherical transition of seismic
  attenuation at the top of the Earth's inner core}. Earth planet. Sci. Lett.
  228, 243--253.

\bibitem[{Chalmers(1964)}]{Chalmers1964}
Chalmers, B., 1964. {Principles of Solidification}. J. Wiley \& Sons, NY,
  319pp.

\bibitem[{Christensen and Tilgner(2004)}]{Christensen2004}
Christensen, U., Tilgner, A., 2004. {Power requirement of the geodynamo from
  ohmic losses in numerical and laboratory dynamos}. Nature 429, 169--171.

\bibitem[{Coriell et~al.(1976)Coriell, Hurle, and Sekerka}]{Coriell76}
Coriell, S., Hurle, D., Sekerka, R., 1976. {Interface stability during crystal
  growth: the effect of stirring}. J. Crystal Growth 32, 1--7.

\bibitem[{{Davis}(2001)}]{Davis01}
{Davis}, S.~H., 2001. {Theory of Solidification}. pp.~400.~Cambridge, UK:
  Cambridge University Press.

\bibitem[{{Dziewonski} and {Anderson}(1981)}]{PREM}
{Dziewonski}, A.~M., {Anderson}, D.~L., 1981. {Preliminary reference Earth
  model}. Phys. Earth Planet. Inter. 25, 297--356.

\bibitem[{Favier and Rouzaud(1983)}]{Favier83}
Favier, J.~J., Rouzaud, A., 1983. {Morphological stability of the
  solidification interface under convective conditions}. J. Crystal Growth 64,
  367--379.

\bibitem[{Fearn et~al.(1981)Fearn, Loper, and Roberts}]{Fearn1981}
Fearn, D., Loper, D., Roberts, P., 1981. {Structure of the Earth's inner core}.
  Nature 292, 232--233.

\bibitem[{{Gubbins} et~al.(2004){Gubbins}, {Alf{\`e}}, {Masters}, {Price}, and
  {Gillan}}]{Gubbins04}
{Gubbins}, D., {Alf{\`e}}, D., {Masters}, G., {Price}, G.~D., {Gillan}, M.,
  2004. {Gross thermodynamics of two-component core convection}. Geophys. J.
  Int. 157, 1407--1414.

\bibitem[{Hunt(1979)}]{Hunt79}
Hunt, J., 1979. Solidification and Casting of Metals. The Metal Society,
  London.

\bibitem[{Jacobs(1953)}]{Jacobs1953}
Jacobs, J., 1953. {The Earth's inner core}. Nature 172, 297.

\bibitem[{{Knittle} and {Jeanloz}(1991)}]{Knittle1991}
{Knittle}, E., {Jeanloz}, R., 1991. {The high-pressure phase diagram of
  Fe(0.94)O - A possible constituent of the earth's core}. J. Geophys. Res. 96,
  16169--80.

\bibitem[{Kurz and Fisher(1989)}]{Kurz89}
Kurz, W., Fisher, D., 1989. {Fundamentals of Solidification}. Trans Tech
  Publications, Switzerland, 305pp.

\bibitem[{{Labrosse} et~al.(2001){Labrosse}, {Poirier}, and {Le
  Mou{\"e}l}}]{Labrosse2001}
{Labrosse}, S., {Poirier}, J.-P., {Le Mou{\"e}l}, J.-L., 2001. {The age of the
  inner core}. Earth planet. Sci. Lett. 190, 111--123.

\bibitem[{Laio et~al.(2000)Laio, Bernard, Chiarotti, Scandalo, and
  Tosatti}]{Laio2000}
Laio, A., Bernard, S., Chiarotti, G.~L., Scandalo, S., Tosatti, E., 2000.
  {Physics of Iron at Earth's Core Conditions}. Science 287, 1027.

\bibitem[{Langer and M{\"u}ller-Krumbhaar(1977)}]{Langer77}
Langer, J., M{\"u}ller-Krumbhaar, H., 1977. {Stability effects in dendritic
  crystal growth}. J. Crystal Growth 42, 11--14.

\bibitem[{Le~Bars and Worster(2006)}]{LeBars06}
Le~Bars, M., Worster, M., 2006. {Interfacial conditions between a pure fluid
  and a porous medium: implications for binary alloy}. J. Fluid Mech. 550,
  149--173.

\bibitem[{Loper and Roberts(1981)}]{Loper1981}
Loper, D., Roberts, P., 1981. {A study of conditions at the inner core boundary
  of the Earth}. Phys. Earth Planet. Inter. 24, 302--307.

\bibitem[{{Loper}(1983)}]{Loper1983}
{Loper}, D.~E., 1983. {Structure of the inner core boundary}. Geophys. and
  Astrophys. Fluid Dyn. 25, 139--155.

\bibitem[{{Morse}(1986)}]{Morse1986}
{Morse}, S.~A., 1986. {Adcumulus growth of the inner core}. Geophys. Res. Lett.
  13, 1466--1469.

\bibitem[{{Morse}(2002)}]{Morse2002}
{Morse}, S.~A., 2002. {No mushy zones in the Earth's core}. Geochimica and
  Cosmochimica Acta 66, 2155--2165.

\bibitem[{Mullins and Sekerka(1963)}]{MS1963}
Mullins, W., Sekerka, R., 1963. {Stability of a planar interface during
  solidification of a dilute binary alloy}. J. Appl. Phys. 35, 444--451.

\bibitem[{{Nimmo} et~al.(2004){Nimmo}, {Price}, {Brodholt}, and
  {Gubbins}}]{Nimmo2004}
{Nimmo}, F., {Price}, G.~D., {Brodholt}, J., {Gubbins}, D., 2004. {The
  influence of potassium on core and geodynamo evolution}. Geophys. J. Int.
  156, 363--376.

\bibitem[{{Poirier}(1994{\natexlab{a}})}]{Poirier1994}
{Poirier}, J.-P., 1994{\natexlab{a}}. {Light elements in the Earth's outer
  core: A critical review}. Phys. Earth Planet. Inter. 85, 319--337.

\bibitem[{{Poirier}(1994{\natexlab{b}})}]{Poirier1994a}
{Poirier}, J.-P., 1994{\natexlab{b}}. {Physical properties of the Earth's
  core}. C.R. Acad. Sci. Paris 318, 341--350.

\bibitem[{{Ringwood} and {Hibberson}(1991)}]{Ringwood1991}
{Ringwood}, A.~E., {Hibberson}, W., 1991. {Solubilities of mantle oxides in
  molten iron at high pressures and temperatures - Implications for the
  composition and formation of Earth's core}. Earth planet. Sci. Lett. 102,
  235--251.

\bibitem[{Rubie et~al.(2004)Rubie, K., and Frost}]{Rubie2004}
Rubie, D.~C., K., G.~C., Frost, D.~J., 2004. {Partitioning of oxygen during
  core formation on the Earth and Mars}. Nature 429, 58--61.

\bibitem[{{Sherman}(1995)}]{Sherman1995}
{Sherman}, D.~M., 1995. {Stability of possible Fe-FeS and Fe-FeO alloy phases
  at high pressure and the composition of the Earth's core}. Earth planet. Sci.
  Lett. 132, 87--98.

\bibitem[{Shimizu et~al.(2005)Shimizu, Poirier, and Le~Mou{\"e}l}]{Shimizu2005}
Shimizu, H., Poirier, J.-P., Le~Mou{\"e}l, J.-L., 2005. {On crystallization at
  the inner core boundary}. Phys. Earth Planet. Inter. 151, 37--51.

\bibitem[{{Stacey} and {Anderson}(2001)}]{Stacey}
{Stacey}, F.~D., {Anderson}, O.~L., 2001. {Electrical and thermal
  conductivities of Fe-Ni-Si alloy under core conditions}. Phys. Earth Planet.
  Inter. 124, 153--162.

\bibitem[{{Stevenson}(1981)}]{Stevenson1981}
{Stevenson}, D.~J., 1981. {Models of the earth's core}. Science 214, 611--619.

\bibitem[{{Stixrude} et~al.(1997){Stixrude}, {Wasserman}, and
  {Cohen}}]{Stixrude1997}
{Stixrude}, L., {Wasserman}, E., {Cohen}, R.~E., 1997. {Composition and
  temperature of Earth's inner core}. J. Geophys. Res. 102, 24729--24740.

\bibitem[{Sumita et~al.(1996)Sumita, Yoshida, Kumazawa, and Hamano}]{Sumita96}
Sumita, I., Yoshida, S., Kumazawa, M., Hamano, Y., 1996. {A model for
  sedimentary compaction of a viscous media and its application to inner-core
  growth}. Geophys. J. Int. 124, 302--324.

\bibitem[{Trivedi(1984)}]{Trivedi84}
Trivedi, R., 1984. {Interdendritic Spacing. II.--A Comparison of Theory and
  Experiment}. Metall. Trans. A. 15A, 977--982.

\bibitem[{{Vo{\v c}adlo} et~al.(2003{\natexlab{a}}){Vo{\v c}adlo}, {Alf{\`e}},
  {Gillan}, and {Price}}]{Vocadlo2003}
{Vo{\v c}adlo}, L., {Alf{\`e}}, D., {Gillan}, M.~J., {Price}, G.~D.,
  2003{\natexlab{a}}. {The properties of iron under core conditions from first
  principles calculations}. Phys. Earth Planet. Inter. 140, 101--125.

\bibitem[{{Vo{\v c}adlo} et~al.(2003{\natexlab{b}}){Vo{\v c}adlo}, {Alf{\`e}},
  {Gillan}, Wood, Brodholt, and {Price}}]{Vocadlo2003b}
{Vo{\v c}adlo}, L., {Alf{\`e}}, D., {Gillan}, M.~J., Wood, I., Brodholt, J.,
  {Price}, G.~D., 2003{\natexlab{b}}. {Possible thermal and chemical
  stabilization of body-centred-cubic iron in the Earth's core}. Nature 424,
  536--539.

\bibitem[{{Wen}(2006)}]{Wen06}
{Wen}, L., 2006. {Localized temporal changes of the Earth's inner core
  boundary}. Science 314, 967--970.

\bibitem[{{Williams} and {Jeanloz}(1990)}]{Williams1990}
{Williams}, Q., {Jeanloz}, R., 1990. {Melting relations in the iron-sulfur
  system at ultra-high pressures - Implications for the thermal state of the
  earth}. J. Geophys. Res. 95, 19299--19310.

\bibitem[{{Yoshida} et~al.(1996){Yoshida}, {Sumita}, and
  {Kumazawa}}]{Yoshida1996}
{Yoshida}, S., {Sumita}, I., {Kumazawa}, M., 1996. {Growth model of the inner
  core coupled with the outer core dynamics and the resulting elastic
  anisotropy}. J. Geophys. Res. 101, 28085--28104.

\bibitem[{{Yukutake}(1998)}]{Yukutake1998}
{Yukutake}, T., 1998. {Implausibility of thermal convection in the Earth's
  solid inner core}. Phys. Earth Planet. Inter. 108, 1--13.

\end{thebibliography}

\end{document}